\documentstyle[12pt]{article}
\newcommand{\mysection}[1]{\setcounter{equation}{0}\section{#1}}

\def\be{\begin{equation}}\def\ee{\end{equation}}\def\l{\label}

\makeatletter\ifcase\@ptsize\font\teneufm=eufm10
\font\seveneufm=eufm7\font\fiveeufm=eufm5
\font\teneusm=eusm10\font\seveneusm=eusm7
\font\fiveeusm=eusm5\or\font\teneufm=eufm10 scaled
\magstephalf\font\seveneufm=eufm7\font\fiveeufm=eufm5
\font\teneusm=eusm10 scaled\magstephalf
\font\seveneusm=eusm7\font\fiveeusm=eusm5\or
\font\teneufm=eufm10 scaled\magstep1\font\seveneufm=eufm7
\font\fiveeufm=eufm5\font\teneusm=eusm10 scaled\magstep1
\font\seveneusm=eusm7\font\fiveeusm=eusm5\fi

\newfam\eufmfam\newfam\eusmfam\textfont\eufmfam=\teneufm
\scriptfont\eufmfam=\seveneufm
\scriptscriptfont\eufmfam=\fiveeufm
\textfont\eusmfam=\teneusm\scriptfont\eusmfam=\seveneusm
\scriptscriptfont\eusmfam=\fiveeusm

\def\frak{\ifmmode\let\next\frak@\else
\def\next{\errmessage{Use\string\frak\space only in math
mode}}\fi\next}\def\frak@#1{{\frak@@{#1}}}
\def\frak@@#1{\fam\eufmfam#1}
\def\sh{\ifmmode\let\next\sh@\else
\def\next{\errmessage{Use\string\sh\space only in math
mode}}\fi\next}\def\sh@#1{{\sh@@{#1}}}
\def\sh@@#1{\fam\eusmfam#1}

\ifcase\@ptsize\font\tenmsa=msam10\font\sevenmsa=msam7
\font\fivemsa=msam5\font\tenmsb=msbm10
\font\sevenmsb=msbm7\font\fivemsb=msbm5\or
\font\tenmsa=msam10 scaled\magstephalf
\font\sevenmsa=msam7\font\fivemsa=msam5
\font\tenmsb=msbm10 scaled\magstephalf
\font\sevenmsb=msbm7\font\fivemsb=msbm5\or
\font\tenmsa=msam10 scaled\magstep1\font\sevenmsa=msam7
\font\fivemsa=msam5\font\tenmsb=msbm10 scaled\magstep1
\font\sevenmsb=msbm7\font\fivemsb=msbm5\fi

\newfam\msafam\newfam\msbfam\textfont\msafam=\tenmsa
\scriptfont\msafam=\sevenmsa
\scriptscriptfont\msafam=\fivemsa\textfont\msbfam=\tenmsb
\scriptfont\msbfam=\sevenmsb
\scriptscriptfont\msbfam=\fivemsb

\def\Bbb{\ifmmode\let\next\Bbb@\else
\def\next{\errmessage{Use\string\Bbb\space only in math
mode}}\fi\next}\def\Bbb@#1{{\Bbb@@{#1}}}
\def\Bbb@@#1{\fam\msbfam#1}\def\hexnumber@#1{\ifnum#1<10
\number#1\else\ifnum#1=10 A\else\ifnum#1=11
B\else\ifnum#1=12 C\else\ifnum#1=13 D\else\ifnum#1=14
E\else\ifnum#1=15 F\fi\fi\fi\fi\fi\fi\fi}
\def\msa@{\hexnumber@\msafam}\def\msb@{\hexnumber@\msbfam}
\mathchardef\square="0\msa@03

\makeatother
\newcommand{\HH}{{\Bbb H}}\newcommand{\RR}{{\Bbb R}}
\newcommand{\CC}{{\Bbb C}}

\newcommand{\II}{{\Bbb I}}

\newcommand{\ba}{\begin{array}}
\newcommand{\ea}{\end{array}}
\newcommand{\bea}{\begin{eqnarray}}
\newcommand{\eea}{\end{eqnarray}}

\begin{document}

\begin{titlepage}

{\hfill MIT-CTP-2957}

{\hfill DFPD99/TH/11}

{\hfill US-FT/4-00}

{\hfill hep-th/0003131}

\begin{center}

\vspace{.333cm}

\centerline{\Large\bf Noncommutative Riemann Surfaces}


\vspace{.999cm}{\centerline{\sc Gaetano BERTOLDI$^{1}$,
Jos\'e M. ISIDRO$^{2}$, Marco MATONE$^{2}$ and Paolo PASTI$^{2}$}}
\vspace{.2in}
{\it $^{1}$ Center for Theoretical Physics\\
    Laboratory for Nuclear Science\\
    and Department of Physics\\
    Massachusetts Institute of Technology, Cambridge, MA 02139, USA\\
e-mail: bertoldi@ctp.mit.edu\\}
\vspace{.08in}
{\it $^{2}$ Dipartimento di Fisica ``G. Galilei'' - Istituto Nazionale di
Fisica Nucleare\\}

{\it Universit\`a di Padova\\}

{\it Via Marzolo, 8 - 35131 Padova, Italy\\
e-mail: isidro, matone, pasti@pd.infn.it\\}
\end{center}
\vspace{1cm}

\centerline{\sc ABSTRACT}

\vspace{0.6cm}

\noindent
We compactify M(atrix)
theory on Riemann surfaces $\Sigma$ with genus $g>1$. 
Following [1], we construct a projective unitary representation of
$\pi_1(\Sigma)$ realized on 
$L^2({\HH})$,
with ${\HH}$ the upper half--plane.
As a first step we introduce a suitably gauged ${\rm sl}_2({\RR})$ algebra.
Then a uniquely determined gauge connection provides
the central extension which is
a 2--cocycle of the 2nd Hochschild cohomology group.
Our construction is the
double--scaling limit $N\to\infty$, $k\to-\infty$
of the representation considered in the
Narasimhan--Seshadri theorem, which represents the higher--genus
analog of 't Hooft's clock and shift matrices of QCD. The concept of
a noncommutative Riemann surface $\Sigma_{\theta}$ is introduced
as a certain $C^\star$--algebra. Finally we investigate
the Morita equivalence.

{\it Contribution to the TMR meeting ''Quantum Aspects of Gauge Theories, Supersymmetry
and Unification", Paris, September 1--7, 1997.}

\end{titlepage}

\newpage

\setcounter{footnote}{0}

\renewcommand{\thefootnote}{\arabic{footnote}}

\mysection{Introduction}

The $P_-=N/R$ sector of the discrete light--cone quantization 
of uncompactified M--theory is
given by the supersymmetric quantum mechanics of $U(N)$ matrices.
The compactification of M(atrix) theory \cite{MATRIX}--\cite{SUSS} 
as a model for M--theory
\cite{REVTAYLOR} has been studied in \cite{TAYLOR}. 
In \cite{CDS}--\cite{CASALBUONI} it has been
treated using noncommutative geometry \cite{CONNESETAL}. 
These investigations apply to the $d$--dimensional torus $T^d$, 
and have been further dealt with from
various viewpoints in \cite{DOUGLAS}--\cite{IRAN}. These structures
are also relevant in noncommutative string and gauge theories
\cite{SWn,SchCorSch}. In this paper, following \cite{LAVORO}, we address the
compactification M(atrix) theory on Riemann surfaces with genus $g>1$.

A Riemann surface $\Sigma$ of genus $g>1$ is constructed 
as the quotient $\HH/\Gamma$, where
$\HH$ is the upper half--plane,
and $\Gamma\subset{\rm PSL}_2({\RR})$,
$\Gamma\cong\pi_1(\Sigma)$, is a Fuchsian
group acting on ${\HH}$ as
\begin{equation} 
\gamma=\left(\begin{array}{c}a\\c
\end{array}\begin{array}{cc}b\\d\end{array}\right)\in\Gamma,\;\qquad
\gamma z={az+b\over cz+d}.
\label{3}
\end{equation}
In the absence of elliptic and
parabolic generators, the $2g$ Fuchsian generators $\gamma_j$  
satisfy
\begin{equation}
\prod_{j=1}^g\left(\gamma_{2j-1}\gamma_{2j}{\gamma_{2j-1}^{-1}}
{\gamma_{2j}^{-1}}\right)={\II}.
\label{3xpergamma}
\end{equation}

Inspired by M(atrix) theory, let us promote the complex coordinate 
$z =x+iy$ to an
$N\times N$ complex matrix $Z=X+iY$, with $X=X^{\dagger}$ and $Y= Y^{\dagger}$.
This suggests defining fractional linear transformations of $Z$ through 
conjugation 
with some non--singular matrix ${\cal U}$:
\begin{equation}
{\cal U} Z{\cal U}^{-1} =(aZ+b {\II})(cZ+d{\II})^{-1}.
\label{moobb2}
\end{equation}
Accordingly, operators ${\cal U}_k$ representing the Fuchsian generators 
$\gamma_k$ can be
constructed, such that
\begin{equation}
\prod_{k=1}^g\left({\cal U}_{2k-1}\,{\cal U}_{2k}\,{\cal U}_{2k-1}^{-1}\,
{\cal U}_{2k}^{-1}\right)=e^{2\pi i\theta}{\II}.
\label{inonni}
\end{equation} 
While we will find the solution to (\ref{inonni}), we will consider
slightly different versions of (\ref{moobb2}).
This construction cannot be implemented for finite $N$, as taking the trace of
(\ref{moobb2}) shows. It can be interpreted as defining a sort of
{\it M(atrix) uniformization}, in which the M\"obius transformation of the
M(atrix) coordinate $Z$ is defined through (\ref{moobb2}). 

\mysection{Compactification in $g>1$}

Next we present an explicit Ansatz to compactify 11--dimensional 
supergravity on a Riemann surface
with $g>1$. The Einstein equations read
$$  
R_{MN}-{1\over 2}G_{MN}R
$$
\be
={1\over 3}(H_{ML_1L_2L_3}H_{NL'_1L'_2L'_3}G^{L_1L'_1}G^{L_2L'_2}G^{L_3L'_3}
-{1\over 8}G_{MN}H_{L_1L_2L_3L_4}H_{L'_1L'_2L'_3L'_4}
G^{L_1L'_1}G^{L_2L'_2} G^{L_3L'_3}G^{L_4L'_4}),
\label{einstein}
\end{equation}  
where $H_{MNPQ}$ is the field strength of $C_{MNP}$. We try an Ansatz
by diagonally decomposing $G_{MN}$ into 2--, 4-- and 5--dimensional blocks, 
with $H_{MNPQ}$ taken along the 4--dimensional subspace:
$$  
G_{MN}={\rm diag}\, (g_{\alpha\beta}^{(2)},g_{mn}^{(4)},g_{ab}^{(5)}),
$$
\begin{equation}  
H_{MPQR}=\epsilon_{mpqr} f.
\label{ansatz}
\end{equation}  
The Einstein equations then decompose as
\be
R^{(k)}_{i_kj_k}-{1\over 2}g^{(k)}_{i_kj_k}(R^{(2)}+R^{(4)}+R^{(5)})  
=\epsilon_k{\rm det}\, g^{(4)}\, f^2\, g^{(k)}_{i_kj_k},
\label{ein}
\end{equation}
where $k=2,4,5$, $(i_2,j_2)=(\alpha,\beta)$, $(i_4,j_4)=(m,n)$, 
$(i_5,j_5)=(a,b)$, and
$\epsilon_2=\epsilon_4=-\epsilon_5=1$.
Some manipulations lead to 
\begin{equation}  
R^{(k)}=c_kf^2 {\rm det}\, g^{(4)},
\label{rscalar}
\end{equation} 
with $c_2=-4/3$, $c_4=16/3$ and $c_5=-10/3$. 
We observe that $f=0$ would reproduce the toroidal case.
A non--vanishing $f$ is a deformation 
producing $g>1$. 
 It suffices that $g^{(4)}$ 
have positive signature for $R^{(2)}$ to be negative,
as required in $g>1$. Then a choice
for the 4-- and 5--dimensional manifolds is $S^4$ and $AdS^5$.

\mysection{Differential representation of $\Gamma$}

\subsection{The unitary gauged operators}

For $n=-1,0,1$ and $e_n(z)=z^{n+1}$ we consider the 
${\rm sl}_2(\RR)$ operators
$\ell_n=e_n(z)\partial_z$. We define 
\begin{equation} 
L_n=e_n^{-1/2}\ell_ne^{1/2}_n= e_n\left(\partial_z+{1\over2}
{e'_n\over e_n}\right).
\label{glielleenne}
\end{equation} 
These satisfy
$$
[L_m,L_n]=(n-m)L_{m+n},\quad [\bar L_m,L_n]=0,
$$
\begin{equation} 
[L_n,f]=z^{n+1}\partial_zf.
\label{x9ewixh}
\end{equation}
For $k=1,2,\ldots, 2g$, consider the operators
\begin{equation}
T_k=e^{\lambda_{-1}^{(k)}(L_{-1}+\bar L_{-1})}e^{\lambda_0^{(k)}(L_0+\bar L_0)}
e^{\lambda_{1}^{(k)}(L_{1}+\bar L_{1})},
\label{iVcappa}
\end{equation}
with the $\lambda_n^{(k)}$ picked such that
$T_kzT_k^{-1}=\gamma_kz=(a_kz+b_k)/
(c_kz+d_k)$ so that by (\ref{3xpergamma}) 
\begin{equation}
\prod_{k=1}^g\left( T_{2k-1}T_{2k}T_{2k-1}^{-1}T_{2k}^{-1}
\right)={\II}.
\label{thetees}
\end{equation} 
On $L^2(\HH)$ we have the scalar product
\begin{equation}
\langle\phi|\psi\rangle=\int_\HH d\nu\bar\phi\psi,
\label{prodottoscaler}
\end{equation}
$d\nu(z)=idz\wedge d\bar z/2=dx\wedge dy$.
The $T_k$ provide a unitary representation
of $\Gamma$.

Next consider the {\it gauged} ${\rm sl}_2(\RR)$ operators \cite{LAVORO}
\be
{\cal L}_n^{(F)}=F(z,\bar z) L_n F^{-1}(z,\bar z)
=e_n\left(\partial_z+{1\over 2}{e'_n\over e_n}-
\partial_z \ln F(z,\bar z)\right),
\label{glielleennecal}
\end{equation} 
where $F(z,\bar z)$ is an undetermined phase function,
to be determined later on. The ${\cal L}_n^{(F)}$ also satisfy 
the algebra (\ref{x9ewixh}).
The adjoint of ${\cal L}_n^{(F)}$ is given by
\begin{equation}
{\cal L}_n^{(F)\dagger}=-F\overline{e^{1/2}_n}\partial_{\bar z}
\overline {e^{1/2}_n}F^{-1},
\label{abbastanzabasilare}
\end{equation}
with ${\cal L}_n^{(F)\dagger}=-\bar{\cal L}_n^{(F^{-1})}$. Finally we define
\begin{equation}
\Lambda_n^{(F)}=
{\cal L}_n^{(F)}-{\cal L}_n^{(F)\dagger}={\cal L}_n^{(F)}+ 
\bar{\cal L}_n^{(F^{-1})}.
\label{zx4}
\end{equation} 
The $\Lambda_n^{(F)}$ enjoy the fundamental property that both their chiral 
components are gauged in the same way by the function $F$, that is
\begin{equation}
\Lambda_n^{(F)}=F(L_n+\bar L_n)F^{-1},
\label{zx5}
\end{equation} 
while also satisfying the ${\rm sl}_2(\RR)$ algebra:
$$ 
[\Lambda_m^{(F)},\Lambda_n^{(F)}]=(n-m)\Lambda_{m+n}^{(F)},
$$
\begin{equation}
[\Lambda_n^{(F)},f]=(z^{n+1}\partial_z+{\bar z}^{n+1}\partial_{\bar z})f.
\label{soddisfanosoddisfano}
\end{equation}
It holds that 
\begin{equation} 
e^{\Lambda_n^{(F)}}=Fe^{L_n+\bar L_n}F^{-1},
\label{zx10}
\end{equation} 
which is a unitary operator since $\Lambda_n^{(F)\dagger}=-\Lambda_n^{(F)}$.

Let $b$ be a real number, and $A$ a Hermitean connection 1--form 
to be identified presently. Set
\begin{equation} 
{\cal U}_k=e^{ib\int_z^{\gamma_k z}A}T_k,
\label{loperatoree}
\end{equation}
where the integration contour is taken to be the Poincar\'e geodesic connecting 
$z$ and $\gamma_k z$.
As the gauging functions introduced in (\ref{glielleennecal}) we will 
take the functions
$F_k(z,\bar z)$ that solve the equation
\begin{equation} 
F_kT_kF^{-1}_k=e^{ib\int_z^{\gamma_kz}A}T_k,
\label{loperatoreeeq}
\end{equation}
that is
\begin{equation}
F_k(\gamma_kz, \gamma_k\bar z)=e^{-ib\int_z^{\gamma_k z} A}F_k(z,\bar z).
\label{thatis}
\end{equation}

\subsection{The gauged algebra}

With the choice (\ref{loperatoreeeq}) for $F_k$, (\ref{zx5}) becomes
\be
\Lambda_{n,k}^{(F)}=F_k(L_n+\bar L_n)F^{-1}_k
=z^{n+1}\left(\partial_z+{n+1\over2z}-\partial_z \ln F_k\right)
+{\bar z}^{n+1}\left(\partial_{\bar z}+{n+1\over2{\bar z}}-
\partial_{\bar z} \ln
F_k\right).
\label{zx5bisse}
\ee
The $\Lambda_{n,k}^{(F)}$ satisfy the algebra
$$
[\Lambda_{m,j}^{(F)},\Lambda_{n,k}^{(F)}]
=(n-m)\Lambda_{m+n,j}^{(F)}
+F_k^{-1}| e_n|\Lambda_{n,k}^{(F)}|e_n|^{-1}F_k
F_j^{-1}|e_m|\Lambda_{m,j}^{(F)}|e_m|^{-1}F_j(\ln F_j- \ln F_k),
$$
\be
[\Lambda_{n,k}^{(F)},f]=(z^{n+1}\partial_z+{\bar z}^{n+1}\partial_{\bar z})f.
\label{commutator}
\end{equation}
Upon exponentiating $\Lambda_{n,k}^{(F)}$ one finds
\begin{equation}
{\cal U}_k=e^{\lambda_{-1}^{(k)}\Lambda_{-1,k}^{(F)}}\,e^{\lambda_0^{
(k)}\Lambda_{0,k}^{(F)}}\,e^{\lambda_{1}^{(k)}\Lambda_{1,k}^{(F)}},
\label{zx11}
\end{equation}
that is, the ${\cal U}_k$ are unitary, and
\begin{equation}
{\cal U}_k^{-1}=T_k^{-1}e^{-ib\int_z^{\gamma_kz}A}=
e^{-ib\int_{\gamma_k^{-1}z}^zA}T_k^{-1}.
\label{iosdq899999}
\end{equation}

\subsection{Computing the phase}

It is immediate to see that the ${\cal U}_k$ defined in (\ref{loperatoree}) 
satisfy
(\ref{inonni}) for a certain value of $\theta$:
$$
\prod_{k=1}^g\left({\cal U}_{2k-1}{\cal U}_{2k}{\cal U}_{2k-1}^{\dagger}
{\cal U}_{2k}^{\dagger}\right)
$$
$$ 
=e^{ib\int_z^{\gamma_1z}A}T_1e^{ib\int_z^{\gamma_2z}A}T_2
e^{-ib\int_{\gamma_1^{-1}z}^zA}T_1^{-1}e^{-ib\int_{\gamma_2^{-1}z}^zA}
T_2^{-1}\ldots
$$
$$
=\exp\left[ib\left(\int_z^{\gamma_1z}A+\int_{\gamma_1z}^{\gamma_2\gamma_1z}
A+\int_{\gamma_2\gamma_1z}^{\gamma_1^{-1}\gamma_2\gamma_1z}A
+\int_{\gamma^{-1}_1
\gamma_2\gamma_1z}^{\gamma_2^{-1}\gamma_1^{-1}\gamma_2\gamma_1z}A+
\ldots\right)\right]
\prod_{k=1}^g\left(T_{2k-1}
T_{2k}T_{2k-1}^{-1}T_{2k}^{-1}\right)
$$
\be
=e^{ib\oint_{\partial{\cal F}_z}A},
\label{abbiamox1}
\end{equation} 
where ${\cal F}_z=\{z,\gamma_1 z,\gamma_2\gamma_1z,\gamma_1^{-1}
\gamma_2\gamma_1 z,\ldots\}$
is a fundamental domain for $\Gamma$.
The basepoint $z$, plus the action of the Fuchsian generators on it, 
determine
${\cal F}_z$, as the vertices are joined by geodesics. 

\subsection{Uniqueness of the gauge connection}

For (\ref{abbiamox1}) to provide a projective unitary representation
of $\Gamma$,  $\int_{{\cal F}_z} dA$ should be $z$--independent.
Changing $z$ to
$z'$ can be expressed as $z\to z'=\mu z$ for some 
$\mu\in{\rm PSL}_2({\RR})$. Then
${\cal F}_z\rightarrow {\cal F}_{\mu z}=
\{\mu z,\gamma_1\mu z,\gamma_2\gamma_1 \mu
z,\gamma_1^{-1}\gamma_2 \gamma_1
\mu z,\ldots\}$.
Now consider
${\cal F}_z\rightarrow \mu
{\cal F}_z=
\{\mu z,\mu\gamma_1 z,\mu\gamma_2\gamma_1 z,
\mu\gamma_1^{-1}\gamma_2 \gamma_1 z,\ldots\}$.
The congruence $\mu {\cal F}_z\cong{\cal F}_{\mu z}$ follows from 
two facts: that the
vertices are joined by geodesics, and that ${\rm PSL}_2({\RR})$ maps geodesics
into geodesics. Since $\Gamma$ is defined up to conjugation, 
$\Gamma\to\mu\Gamma\mu^{-1}$, if $\mu{\cal F}_z$ is a fundamental domain, 
so is ${\cal F}_{\mu z}$.  Thus,
to have $z$--independence we need $\forall\mu\in{\rm PSL}_2({\RR})$
\begin{equation}
\int_{{\cal F}_z}dA=\int_{{\cal F}_{\mu z}}dA=\int_{\mu{\cal F}_z}dA
=\int_{{\cal F}}dA.
\label{0ijqI}
\end{equation}
This fixes the
(1,1)--form $dA$ to be ${\rm PSL}_2({\RR})$--invariant.
It is well known that the Poincar\'e form is the unique
${\rm PSL}_2({\RR})$--invariant (1,1)--form, up to an overall constant
factor. This is a particular case of a more general fact \cite{NOI}.
The Poincar\'e metric $ds^2=y^{-2}|dz|^2=2g_{z\bar z}|dz|^2=
e^{\varphi}|dz|^2$ has curvature
$R=-g^{z\bar z}\partial_z\partial_{\bar z}\ln \, g_{z\bar z}=-1$, so that 
$\int_{{\cal F}} d\nu e^{\varphi}= -2\pi\chi(\Sigma)$, 
where $\chi(\Sigma)=2-2g$ is the Euler
characteristic.  As  the Poincar\'e (1,1)--form is 
$dA=e^\varphi d\nu$,
this uniquely determines the gauge field to be 
\begin{equation}    
A=A_zdz+A_{\bar z}d\bar z={dx\over y},
\label{Aconn}
\end{equation}   
up to gauge transformations. Using 
$\oint_{\partial{\cal F}}A=\int_{\cal F} dA$
 we finally have that (\ref{abbiamox1}) becomes
\begin{equation}
\prod_{k=1}^g\left({\cal U}_{2k-1}{\cal U}_{2k}{\cal U}_{2k-1}^{\dagger}
{\cal U}_{2k}^{\dagger}\right)=
e^{2\pi ib\chi(\Sigma)}.
\label{perfect}
\end{equation} 

\subsection{Non--Abelian extension}

Up to now we considered the case in which the connection is
Abelian. However, it is easy to extend our construction to the
non--Abelian case in which the gauge group
$U(1)$ is replaced by
$U(N)$. The operators ${\cal U}_k$ now become
\begin{equation}
{\cal U}_k=P e^{ib\int^{\gamma_k z}_z A} T_k,
\label{cdo3jn}\end{equation}
where the $T_k$ are the same as before, times the $N\times N$ identity matrix.
Eq.(\ref{abbiamox1}) is replaced by
\begin{equation}
\prod_{k=1}^g\left({\cal U}_{2k-1}{\cal U}_{2k}{\cal U}_{2k-1}^{\dagger}
{\cal U}_{2k}^{\dagger}\right)
= P e^{ib\oint_{\partial{\cal F}_z}A}.
\label{abbiamox1nonabeliana}
\end{equation}
Given an integral
along a closed contour $\sigma_z$ with basepoint
$z$, the path--ordered exponentials for
a connection $A$ and its gauge transform
$A^U=U^{-1}AU+U^{-1}dU$ are related by \cite{THOMPSON}
\begin{equation}
P e^{i\oint_{\sigma_z}A}=U(z)
P e^{i\oint_{\sigma_z}A^U}U^{-1}(z)=
U(z)P e^{i\oint_{\sigma_z}d\sigma^\mu\int_0^1ds s\sigma^\nu
U^{-1}(s\sigma)F_{\nu\mu}
(s\sigma)U(s\sigma)}U^{-1}(z).
\label{buonina}\end{equation} 
Applying this to (\ref{abbiamox1nonabeliana}), we see that the only
possibility to get a coordinate--independent phase is for
the curvature (1,1)--form $F=dA+[A,A]/2$
to be the identity matrix in the gauge indices times a (1,1)--form
$\eta$, that is $F=\eta{\II}$.
It follows that
\begin{equation}
P e^{ib\oint_{\partial{\cal F}}A}=
e^{ib\int_{{\cal F}}F}.
\label{boh}\end{equation}
However, the above is only a necessary
condition for coordinate--independence. Nevertheless,
we can apply the same reasoning as in the Abelian case
to see that $\eta$ should be
proportional to the Poincar\'e (1,1)--form.
Denoting by $E$ the vector bundle on which $A$ is defined, 
we have $k={\rm deg}\,(E)={1\over 2\pi}{\rm tr}\,\int_{{\cal F}} F$.
Set $\mu(E)=k/N$ so that
$\int_{{\cal F}}F=2\pi \mu(E){\II}$ and
$\eta=-{\mu(E)\over \chi(\Sigma)} e^\varphi d\nu$, {\it i.e.}
\begin{equation}
F=2\pi\mu(E)\omega {\II},
\label{zummmolloo}\end{equation}
where $\omega=\left(e^\varphi/\int_{{\cal F}}d\nu e^\varphi\right)
d\nu$. Thus, by
(\ref{boh}) we have that Eq.(\ref{abbiamox1nonabeliana}) becomes
\begin{equation}
\prod_{k=1}^g\left({\cal U}_{2k-1}{\cal U}_{2k}{\cal U}_{2k-1}^{\dagger}
{\cal U}_{2k}^{\dagger}\right)
= e^{2\pi i b\mu(E)}{\II},
\label{abbiamox1nonabelianaduino}
\end{equation}
which provides a projective unitary representation
of $\pi_1(\Sigma)$ on $L^2(\HH,\CC^N)$.

\subsection{The gauge length}

A basic object is the {\it gauge length} function
\begin{equation}   
d_{A}(z,w)=\int^w_zA,
\label{elleee}
\end{equation}  where the contour integral is along
the Poincar\'e geodesic connecting $z$ and $w$.
In the Abelian case 
\begin{equation}
d_{A}(z,w)=\int^{{\rm Re}\,w}_{{\rm Re}\,z}{dx\over y}
=-i\ln\left( {z-\bar w\over w-\bar z}\right),
\label{logaritmo}
\end{equation}
which is equal to the angle $\alpha_{zw}$ spanned by the arc of geodesic
connecting $z$ and $w$.
Observe that the gauge length of the geodesic connecting two punctures,
{\it i.e.} two points on the real line, is $\pi$. This is to be compared 
with the usual divergence of the Poincar\'e distance.
Under a ${\rm PSL}_2({\RR})$--transformation $\mu$, we have
($\mu_x\equiv\partial_x \mu x$)
\begin{equation}
d_{A}\left(\mu z,\mu w\right)=d_{A}(z,w)  
-{i\over 2}\ln\left({\mu_z\bar \mu_w\over\bar\mu_z \mu_w}\right).
\label{ioeuIpo}
\end{equation}
Therefore, the gauge length of an $n$--gon
\begin{equation} 
d^{(n)}_{A}(\{z_k\})= \sum_{k=1}^nd_{A}(z_k,z_{k+1})
=\pi(n-2)-\sum_{k=1}^n\alpha_k,
\label{poli4w}
\end{equation}
where $z_{n+1}\equiv z_1$, $n\geq 3$, and $\alpha_k$
are the internal angles,
is ${\rm PSL}_2({\RR})$--invariant.
One can check that the
${\rm PSL}_2({\RR})$--transformation (\ref{ioeuIpo}) corresponds
to a gauge transformation of $A$. Furthermore, as we will see,
the triangle length,
that by Stokes' theorem corresponds to the Poincar\'e area,
is proportional to the Hochschild 2--cocycle.

\subsection{Pre--automorphic forms}

A related reason for the relevance of
the gauge length function is that it also appears in the
definition of the $F_k$. The latter, which apparently
never appeared in the literature before, are
of particular interest.
Let us recast (\ref{loperatoreeeq}) as
\begin{equation}
F_k(\gamma_k z,\gamma_k\bar z)=\left({\gamma_k z-\bar z\over 
z-\gamma_k \bar z}\right)^b\,F_k(z,\bar z).
\label{effecapppa4}
\end{equation}
Since $(\gamma_k z-\bar z)/(z-\gamma_k \bar z)$
transforms as an automorphic form under $\Gamma$,
we call the $F_k$ {\it pre--automorphic forms}.
Eq.(\ref{thatis}) indicates that finding the most general solution to 
(\ref{effecapppa4}) is
a problem in geodesic analysis. In the case of the inversion
$\gamma_kz=-1/z$ and $b$ an even integer, a solution to 
(\ref{effecapppa4}) is 
$F_k=\left(z/\bar z\right)^{b\over2}$.  By (\ref{logaritmo})
$F_k=\left(z/ \bar z\right)^{b\over2}$ is related to the {\it $A$--length} of
the geodesic connecting $z$ and $0$:
\begin{equation} 
e^{{i\over2}b\int_z^0A}=F_k(z,\bar z)=\left({z\over\bar z}\right)^{b\over2}.
\label{idwjhyu}
\end{equation}
An interesting formal solution to (\ref{effecapppa4}) is
\begin{equation}
F_k(z,\bar z)=\prod_{j=0}^\infty\left({\gamma_k^{-j}z-\gamma_k^{-j-1} 
\bar z\over
\gamma_k^{-j-1} z-\gamma_k^{-j} \bar z}\right)^b.
\label{pro}
\end{equation}
To construct other solutions, we consider 
the uniformizing map
$J_{\HH}:{\HH}\longrightarrow\Sigma$,
which enjoys the property $J_{\HH}(\gamma z)=J_{\HH}(z)$,
$\forall\gamma\in\Gamma$. Then, 
if $F_k$ satisfies (\ref{effecapppa4}), this equation is invariant under
$F_k\rightarrow G(J_{\HH},\bar J_{\HH})F_k$.
Since $|F_k|=1$,
we should require $|G|=1$, otherwise $G$ is arbitrary.

\mysection{Hochschild cohomology of $\Gamma$}

The Fuchsian generators $\gamma_k\in\Gamma$ are projectively represented 
by means of unitary
operators ${\cal U}_k$ acting  on $L^2({\HH})$. The
product $\gamma_k\gamma_j$ is represented
by\footnote{The differential representation of ${\rm PSL}_2({\RR})$
acts in reverse order with respect to the one by matrices.}
${\cal U}_{jk}$, which equals ${\cal U}_j{\cal U}_k$ up to a phase:
\begin{equation}
{\cal U}_j{\cal U}_k=e^{2\pi i\theta(j,k)}{\cal U}_{jk}.
\label{projectivity}
\end{equation}
Associativity implies
\begin{equation}
\theta(j,k) + \theta(jk,l)=
\theta(j,kl) + \theta(k,l).
\label{cocycle}
\end{equation}
We can easily determine $\theta(j,k)$:
\be
{\cal U}_j{\cal U}_k=\exp\left(ib\int_z^{\gamma_j
z}A+ib\int_{\gamma_jz}^{\gamma_k\gamma_jz}A
-ib\int_z^{\gamma_k\gamma_jz}A\right){\cal U}_{jk} =
\exp\left(ib\int_{\tau_{jk}}A\right){\cal U}_{jk},
\label{coc}
\end{equation}
where $\tau_{jk}$ denotes the geodesic triangle with vertices $z$, 
$\gamma_jz$ and
$\gamma_k\gamma_jz$. This identifies $\theta(j,k)$ as the gauge
length of the perimeter of the
geodesic triangle $\tau_{jk}$. By Stokes' theorem this is
the Poincar\'e area of the triangle.
A similar phase, introduced
independently of any gauge connection, has been considered in
\cite{RADULESCU} in the
context of Berezin's quantization of ${\HH}$ and Von Neumann algebras.

The information on the compactification of M(atrix) theory is encoded in
the action of $\Gamma$
on ${\HH}$, plus a projective representation of $\Gamma$. 
The latter amounts to the choice
of a phase. Physically inequivalent choices of $\theta(j,k)$ turn out to be 
in one--to--one
correspondence with elements in the 2nd Hochschild cohomology group 
$H^2(\Gamma, U(1))$ of $\Gamma$. This cohomology group is defined as follows. 
A $k$--cochain is an angular--valued function
$f(\gamma_1,\ldots,\gamma_k)$ with $k$ arguments in $\Gamma$. The coboundary
operator $\delta$ maps the $k$--cochain $f$ into the $(k+1)$--cochain 
$\delta f$ defined as
\be
(\delta f)(\gamma_0,\dots,\gamma_k)=f(\gamma_1,\ldots,\gamma_k)
+\sum_{l=1}^k(-1)^lf(\gamma_0,\ldots,\gamma_{l-1}\gamma_l,\ldots,\gamma_k)
+(-1)^{k+1}f(\gamma_0,\ldots,\gamma_{k-1}).
\label{cochain}
\end{equation}
Clearly $\delta^2=0$. A $k$--cochain annihilated by $\delta$ is called a
$k$--cocycle. $H^k(\Gamma,U(1))$ is the group of equivalence classes of
$k$--cocycles modulo the coboundary of $(k-1)$--cochains. The associativity 
condition (\ref{cocycle})
is just $\delta\theta(j,k)=0$. Thus $\theta$ is a 2--cocycle of the Hochschild
cohomology. Projective representations of $\Gamma$ are classified by
$H^2(\Gamma, U(1))=U(1)$. Hence $\theta=b\chi(\Sigma)$ is the unique
parameter for this compactification ($\theta=b\mu(E)$ in the general case).

\mysection{Stable bundles and double scaling limit} 

We now present some facts about
projective, unitary representations of
$\Gamma$ and the theory of holomorphic vector bundles
\cite{KOBAYASHI,AtiyahSenGuptaFine}.
Let $E\rightarrow \Sigma$ 
be a holomorphic vector bundle over $\Sigma$ of rank $N$ and degree $k$.
The bundle $E$ is called {\it stable}   
if the inequality $\mu(E')<\mu(E)$ 
holds for every proper holomorphic subbundle $E'\subset E$.  
We may take $-N<k\leq 0$.  We will further assume that 
$\Gamma$ contains a unique primitive elliptic 
element $\gamma_0$ of  order $N$ ($i.e.$, $\gamma_0^N={\II}$), 
with fixed point $z_0\in{\HH}$ that projects  to $x_0\in\Sigma$.

Given the branching order $N$ of $\gamma_0$, 
let $\rho:\Gamma\to U(N)$ be an irreducible unitary representation.
It is said  {\it admissible} if 
$\rho(\gamma_0)=e^{-2\pi i k/N}{\II}$.
Putting the elliptic element on the right--hand side, and
setting $\rho_k\equiv\rho(\gamma_k)$,
(\ref{3xpergamma}) becomes
\begin{equation} 
\prod_{j=1}^g\left(\rho_{2j-1}\rho_{2j}
\rho_{2j-1}^{-1}\rho_{2j}^{-1}\right)=e^{2\pi i k/N}{\II}. 
\label{reprho} 
\end{equation} 

On the trivial bundle 
${\HH}\times {\CC}^N\rightarrow {\HH}$ 
there is an action of $\Gamma$: 
$(z, v)\to(\gamma z, \rho(\gamma)v)$. 
This defines the quotient bundle 
\begin{equation} 
{\HH}\times {\CC}^N/\Gamma\rightarrow {\HH}/\Gamma\cong\Sigma. 
\label{trivialquot} 
\end{equation} 
Any admissible representation determines a holomorphic vector bundle  
$E_{\rho}\rightarrow \Sigma$ of rank $N$ and degree $k$.  
When $k=0$, $E_{\rho}$ is simply the 
quotient bundle (\ref{trivialquot}) of
${\HH}\times {\CC}^N\rightarrow {\HH}$. The
Narasimhan--Seshadri (NS) theorem \cite{NASE}
now states that a holomorphic vector bundle $E$ over $\Sigma$ of rank $N$ and degree $k$ 
is stable if and only if it is isomorphic to a  bundle $E_{\rho}$, where $\rho$ 
is an admissible representation of $\Gamma$.  
Moreover, the  bundles $E_{\rho_1}$ and $E_{\rho_2}$ 
are isomorphic if and only if the representations $\rho_1$  
and $\rho_2$ are equivalent. 
 
The standard Hermitean metric on ${\CC}^N$ gives a metric on
${\HH}\times {\CC}^N\rightarrow {\HH}$. 
This metric and the corresponding connection are invariant with respect to 
the action $(z,v)\to (\gamma z,\rho(\gamma) v)$, when $\rho$ is admissible. 
Hence they determine a (degenerate) metric $g_{NS}$ and 
a connection 
$A_{NS}$ on the bundle $E=E_{\rho}$. The connection $A_{NS}$ is  
compatible with the metric 
$g_{NS}$ 
and with the holomorphic structure on $E$,  
but it has a singularity at the branching point 
$x_0\in 
\Sigma$ of the covering ${\HH}\rightarrow \Sigma$.  
The curvature $F_{NS}$ of $A_{NS}$ is a 
$(1,1)$--current 
with values in the bundle ${\rm 
End}\,E$, 
characterized by the property\footnote{Note that our convention
for $A$
differs from the one in the mathematical literature by a factor $i$.}
\begin{equation} 
\int_{\Sigma}f\wedge F_{NS}=-2\pi i\mu (E) {\rm tr}\,f(x_0), 
\label{property} 
\end{equation}  
for every smooth section $f$ of the bundle ${\rm End}\, E$.  
The connection $A_{NS}$ 
is uniquely determined by the curvature condition (\ref{property})  
and by the fact that it 
corresponds to the degenerate metric $g_{NS}$.  
The connection $A_{NS}$ on the stable bundle 
$E=E_{\rho}$ is called the {\it NS connection}.
 
A differential--geometric approach to stability has been given by  
Donaldson \cite{DONALDSON}. 
Fix a Hermitean metric on $\Sigma$,
for example the Poincar\'e metric,
normalized so that the area of
$\Sigma$ equals 1. Let us denote by $\omega$ its associated
(1,1)--form. A holomorphic 
bundle $E$ is stable if and only if 
there exists on $E$ a metric connection $A_D$ with 
central curvature $F_D=-2\pi i \mu(E) \omega
{\II}$; such a connection $A_D$ is unique.

The unitary projective representations
of $\Gamma$ we constructed above have a uniquely defined gauge field 
whose curvature is proportional to the volume form on $\Sigma$. 
With respect to the representation considered
by NS, we note that NS introduced an elliptic point 
to produce the phase, while in our case the latter arises from the gauge length. 
Our construction is directly connected with Donaldson's approach as
$F=iF_D$, where $F$ is the curvature
(\ref{zummmolloo}). However, the main difference is that
our operators are unitary
differential operators on $L^2({\HH},{\CC}^N)$ instead of 
unitary matrices on ${\CC}^N$.
This allowed us to obtain a non--trivial phase also in the Abelian case.

It is however possible to understand the formal relation between 
our operators and those of NS. To see this we consider
the adjoint representation
of $\Gamma$ on ${\rm End}\,{\CC}^N$,
\begin{equation}  
{\rm Ad}\,\rho (\gamma) Z =\rho (\gamma) Z \rho^{-1}(\gamma), 
\label{aggiunta} 
\end{equation} 
where $Z\in{\rm End}\, {\CC}^N$ is understood as an $N\times N$ matrix.  
Let us also consider the trivial bundle 
${\HH}\times{\rm End}\,{\CC}^N\rightarrow {\HH}$.
There is an action of $\Gamma$:
$(z,Z)\mapsto(\gamma z, {\rm Ad}\,\rho(\gamma) Z)$ 
that defines the quotient bundle 
\be
{\HH}\times {\rm End}\, {\CC}^N/\Gamma\rightarrow {\HH}/ 
\Gamma\cong\Sigma.
\l{quotient2}\ee
Then, the idea
is to consider a vector bundle $E'$ in the
double scaling limit
$N'\to\infty$, $k'\to-\infty$, with $\mu(E')=k'/N'$ fixed, that is
\begin{equation}
\mu(E')=b\mu(E).
\label{fixed}\end{equation}
In this limit,
fixing a basis in $L^2({\HH},\CC^N)$, the matrix elements 
of our operators can be identified with those of $\rho(\gamma)$.

\mysection{Noncommutative Riemann surfaces}

Let us now introduce two copies of the upper half--plane, one with
coordinates $z$ and $\bar z$, the other with coordinates $w$ and
$\bar w$. While the coordinates $z$ and $\bar z$ are reserved to the
operators ${\cal U}_k$ we introduced previously, we reserve $w$ and $\bar w$
to construct a new set of operators. We now introduce noncommutative
coordinates expressed in terms of the covariant derivatives
\begin{equation}
W=\partial_w+iA_w, \qquad \bar W=\partial_{\bar w}+iA_{\bar w},
\label{curvattture2}\end{equation}
with $A_w=A_{\bar w}=1/(2\,{\rm Im}\, w)$, so that
\begin{equation}
[W,\bar W]=iF_{w\bar w},
\label{curvattture}\end{equation}
where $F_{w\bar w}=i/[2({\rm Im}\, w)^2]$.
Let us consider
the following realization of the
${\rm sl}_2({\RR})$ algebra:
\begin{equation}
\hat L_{-1}=-w,\qquad \hat L_0=-{1\over 2}(w\partial_w+\partial_w w),
\qquad
\hat L_1=-\partial_w w\partial_w.
\label{sveglio}\end{equation}
We then define the unitary operators
\begin{equation}
\hat T_k=e^{\lambda_{-1}^{(k)}(\hat L_{-1}+\bar{\hat L}_{-1})}
e^{\lambda_0^{(k)}(\hat L_0+\bar{\hat L}_0)}
e^{\lambda_{1}^{(k)}(\hat L_{1}+\bar{\hat L}_{1})},
\label{iVcappatree}
\end{equation}
where the $\lambda_n^{(k)}$ are as in (\ref{iVcappa}).
Set ${\cal V}_k=\hat T_k{\cal U}_k$. Since
the $\hat T_k$ satisfy (\ref{thetees}), it follows that the
${\cal V}_k$ satisfy (\ref{abbiamox1nonabelianaduino}) and
\begin{equation}
{\cal V}_k \partial_w {\cal V}_k^{-1}=
\hat T_k \partial_w {\hat T_k}^{-1}={a_k\partial_w +b_k\over 
c_k\partial_w+d_k}.
\label{dajjiee}\end{equation}
Setting $W=G\partial_wG^{-1}$, {\it i.e.} $G=(w-\bar w)^2$,
and using $Af(B)A^{-1}=f(ABA^{-1})$, we see that
\begin{equation}
{\cal V}_k W {\cal V}_k^{-1}=
\hat T_k W {\hat T_k}^{-1}
=G(\tilde w){\hat T_k}
 \partial_w {\hat T_k}^{-1} G^{-1}(\tilde w),
\label{poidhwd}\end{equation}
where
\begin{equation}
\tilde w={\hat T_k} w {\hat T_k}^{-1}=
-e^{-\lambda_0^{(k)}}+2\lambda_1^{(k)}(\hat L_0-\lambda_{-1}^{(k)}w)
-\lambda_1^{(k)2}e^{\lambda_0^{(k)}}(\hat L_1+2\lambda_{-1}^{(k)}
\hat L_0-{\lambda_{-1}^{(k)2}}w),
\label{robbanova}\end{equation}
and by (\ref{dajjiee})
\begin{equation}
{\cal V}_k W {\cal V}_k^{-1}=
\hat T_k W {\hat T_k}^{-1}={a_k\tilde W +b_k\over 
c_k\tilde W+d_k},
\label{dajjieedue}\end{equation}
where $\tilde W$ differs from $W$ by the connection
\begin{equation}
\tilde W=\partial_w+G(\tilde w)[\partial_w G^{-1}(\tilde w)].
\label{oidhd9oh}\end{equation}

\subsection{{\bf $C^\star$--algebra}}

By a natural generalization of the $n$--dimensional
noncommutative torus, one defines a noncommutative Riemann surface
$\Sigma_\theta$ in $g>1$ to be an associative algebra with involution having
unitary generators ${\cal U}_k$ obeying the relation (\ref{perfect}).
Such an algebra is a $C^\star$--algebra, as it admits a faithful unitary 
representation on $L^2({\HH},{\CC}^N)$ whose image is norm--closed.
Relation (\ref{perfect}) is also satisfied by the ${\cal V}_k$. 
However, while the ${\cal U}_k$ act on the commuting
coordinates $z, \bar z$, the ${\cal V}_k$ act on the operators $W$ and $\bar W$
of (\ref{curvattture2}). The latter, factorized by the action of the ${\cal V}_k$
in (\ref{dajjieedue}), can be pictorially identified with
a sort of noncommutative 
coordinates on $\Sigma_\theta$. 

Each $\gamma\neq {\II}$ in
$\Gamma$ can be uniquely expressed as a positive power of a primitive 
element $p\in\Gamma$, {\it primitive} meaning that $p$ is not a
positive power of any other
$p'\in\Gamma$ \cite{MCKEAN}.
Let ${\cal V}_p$ be the representative of
$p$. Any ${\cal V} \in C^\star$ can 
be written as
\begin{equation}
{\cal V}=\sum_{p\in\{prim\}}\sum_{n=0}^{\infty}c_n^{(p)}{\cal V}_p^n + c_0{\II},
\label{prim}
\end{equation}
for certain coefficients $c_n^{(p)}$, $c_0$. A trace  
can be defined as ${\rm tr}\,{\cal V}=c_0$.

In the case of the torus one can connect the $C^\star$--algebras
of $U(1)$ and $U(N)$. To see this one can use 't Hooft's
clock and shift matrices
\begin{equation}
V_1V_2=e^{2\pi i{M\over N}}V_2V_1.
\label{cazzarola}\end{equation}
The $U(N)$ $C^\star$--algebra 
is constructed in terms of the $V_k$ and of the unitary operators
representing the $U(1)$ $C^\star$--algebra. 
Morita equivalence is an isomorphism
between the two.
In higher genus, the analog of the $V_k$
is the $U(N)$ representation $\rho(\gamma)$
considered above. 
One can obtain a $U(N)$
projective unitary differential representation of $\Gamma$
by taking ${\cal V}_k\rho(\gamma_k)$,
with ${\cal V}_k$ Abelian.
This non--Abelian representation should be compared with
the one obtained by the non--Abelian ${\cal V}_k$
constructed above. In this framework it should be possible to
understand a possible
higher--genus analog of the Morita equivalence.

The isomorphism of the
$C^\star$--algebras is a direct consequence
of an underlying equivalence between the $U(1)$
and $U(N)$ connection. The $z$--independence of the
phase requires $F$ to be the identity matrix in the gauge indices.
This in turn is deeply related to the uniqueness of the
connection we found. The latter is
related to the uniqueness of the NS connection. We conclude
that Morita equivalence in higher genus
is intimately related to the NS theorem.

Finally let us observe that,
as our operators correspond to the $N\to\infty$ limit of
projective unitary representations of $\Gamma$, these play
a role in the $N\to\infty$ limit of QCD
as considered in \cite{BOCHICCHIO}.

\vspace{1cm}

{\bf Acknowledgments.}
It is a pleasure to thank D. Bellisai, D. Bigatti, M. Bochicchio,
U. Bruzzo, R. Casalbuoni, G. Fiore, L. Griguolo, P.M. Ho,
S. Kobayashi, I. Kra, G. Landi, K. Lechner, F. Lizzi, P.A. Marchetti,
B. Maskit, F. R\u adulescu, D. Sorokin, W. Taylor, M. Tonin and
R. Zucchini for comments and interesting discussions. G.B. is supported
in part by  a D.O.E. cooperative agreement DE-FC02-94ER40818 and by an INFN ``Bruno
Rossi" Fellowship. J.M.I. is supported by an INFN fellowship. J.M.I., M.M. and
P.P. are partially supported by the European Commission TMR program
ERBFMRX-CT96-0045.

\end{document}